\documentclass{article}
\usepackage{ijcai11}

\usepackage{times}

\usepackage{amssymb}
\usepackage{graphicx}
\usepackage{subfig}
\usepackage{amsmath}
\usepackage{txfonts}
\usepackage{wrapfig}

\title{Efficient Detection of Hot Span in Information Diffusion from Observation}
\author{Kouzou Ohara$^1$ \quad Kazumi Saito$^2$ \quad Masahiro Kimura$^3$ \quad Hiroshi Motoda$^4$\\
$^1$College of Science and Engineering, Aoyama Gakuin University, ohara@it.aoyama.ac.jp\\
$^2$School of Administration and Informatics, University of Shizuoka, k-saito@u-shizuoka-ken.ac.jp\\
$^3$Department of Electronics and Informatics, Ryukoku University, kimura@rins.ryukoku.ac.jp\\
$^4$Institute of Scientific and Industrial Research, Osaka University, motoda@ar.sanken.osaka-u.ac.jp}

\begin{document}

\maketitle

\vspace{3cm}

\begin{abstract}
We addressed the problem of detecting the change in behavior of
information diffusion from a small amount of observation data, where the
behavior changes were assumed to be effectively reflected in changes in
the diffusion parameter value. The problem is to detect where in time
and how long this change persisted and how big this change is. We solved
this problem by searching the change pattern that maximizes the
likelihood of generating the observed diffusion sequences. The naive
learning algorithm has to iteratively update the patten boundaries, each
requiring optimization of diffusion parameters by the EM algorithm, and
is very inefficient. We devised a very efficient search algorithm using
the derivative of likelihood which avoids parameter value optimization
during the search. The results tested using three real world network
structures confirmed that the algorithm can efficiently identify the
correct change pattern. We further compared our algorithm with the naive
method that finds the best combination of change boundaries by an
exhaustive search through a set of randomly selected boundary
candidates, and showed that the proposed algorithm far outperforms the
native method both in terms of accuracy and computation time.
\end{abstract}

\section{Introduction}\label{intro}

Social networking is now an important part of our daily lives, and our
behavioral patterns are substantially affected by the communication
through these
networks~\cite{newman:physrev02,newman:siam,gruhl:sigkdd,domingos:ieee,leskovec:ec}.
It has been shown that a social network has many interesting
properties, e.g. power law for node degree distribution, large
clustering coefficient, positive degree correlation,
etc.~\cite{wasserman:book94}, which affect how the
information actually diffuses through the network, and researchers have
devised several important measures to characterize these features based
on the topology/structure of the
network~\cite{wasserman:book94,bonacich,katz}.  These measures, called
centrality measures, are expected to be used to identify important nodes
in the network. 
However, recent studies have shown that it is important to consider the
diffusion mechanism explicitly and the measures based on network
structure alone are not enough to identify the important
nodes~\cite{kimura:tkdd,kimura:dmkd}.

Information diffusion is modeled typically by a probabilistic
model. Most representative and fundamental ones are independent cascade
(IC) model~\cite{goldenberg,kempe:kdd}, linear threshold (LT)
model~\cite{watts,watts:ccr} and their extensions that include
incorporating asynchronous time delay~\cite{saito:acml09}.
Explicit use of these models to solve such problems as the {\em
influence maximization problem}~\cite{kempe:kdd,kimura:dmkd} and the
{\em contamination minimization problem}~\cite{kimura:tkdd} clearly
shows the advantage of the model. The identified influential nodes and
links are considerably different from the ones identified by the
centrality measures.  However, use of these models brings in yet another
difficulty. They have parameters that need be specified in advance,
e.g. diffusion probabilities for the IC model, and weights for the LT
model, and their true values are not known in practice.  A series of
studies by \cite{
saito:acml09,
saito:ecml10} have
shown one way of solving this problem in which they used a limited amount
of observed information diffusion data and trained/learned the model
such that the likelihood of generating the observed data by the model is
maximized.

This paper is in the same line of these studies, but addresses a
different aspect of information diffusion. Almost all of the work so far
assumed that the model is stationary. 
We note that our behavior is affected not only by the behaviour of our
neighbors but also by other external factors. The model only accounts for
the interaction with neighbors. 
The problem we address here is to detect the change of the model from a
limited amount of observed information diffusion data. If this is
possible, 
we can infer that something unusual happened during a particular period
of time by simply analyzing the limited amount of data.

This is in some sense the same, in the spirit, with the work by
\cite{kleinberg} and 
~\cite{swan}. They noted a huge volume of the data stream, tried to
organize it and extract structures behind it. This is done in a
retrospective framework, i.e. assuming that there is a flood of abundant
data already and there is a strong need to understand it.
Our aim is not exactly the same as theirs. We are interested in
detecting changes which is hidden in the data. We also follow the same
retrospective approach, i.e. we are not predicting the future, but we
are trying to understand the phenomena that happened in the past. There
are many factors that bring in changes and the model cannot accommodate
all of them. We formalize this as the unknown changes in the
diffusion parameter value, and we reduce the problem to that of
detecting where in time and how long this change persisted and how big
this change is. 
We call the period where the parameter takes anomalous values as
``hot span'' and the rest as ``normal span''. 
To make the analysis simple, we limit the diffusion
model to the asynchronous independent cascade model
(AsIC)~\cite{saito:acml09} and the form of change to a rect-linear one,
that is, the diffusion parameter changes to a new large value, persists
for a certain period of time and is restored to the original value and
stays the same thereafter
\footnote{We discuss that the basic algorithm can be extended to more
general change patterns in Section \ref{discussion}, and shows that it
works for two distinct rect-linear patterns.}.
In this simplified setting, detecting the hot span is equivalent to 
identifying the time window where the parameter value is high and 
estimating the parameter values both in hot and normal spans.

To this end, 
we use the same parameter optimization
algorithm as in \cite{saito:acml09}, i.e. the EM algorithm that
iteratively updates the values to maximize the model's likelihood of
generating the observed data sequences. The problem here is more
difficult because it has another loop to search for the hot span on top
of the above loop. The naive learning algorithm has to iteratively
update the patten boundaries requiring the parameter value optimization
for each combination, which is a very inefficient procedure. Our main
contribution is that we devised a very efficient general search
algorithm which avoids the inner loop optimization by using the
information of the first order derivative of the likelihood with respect
to the diffusion parameters. We tested its performance using the
structures of three real world networks (blog, Coauthorship and
Wikipedia), and confirmed that the algorithm can efficiently identify
the hot span correctly as well as the diffusion parameter values. We
further compared our algorithm with the naive method that finds the best
combination of change boundaries by an exhaustive search from a set of
randomly selected boundary candidates, and showed that the proposed
algorithm far outperforms the native method both in terms of accuracy
and computation time.


\section{Information Diffusion Model}\label{model}

The AsIC model we use in this paper incorporates asynchronous time delay
into the independent cascade (IC) model which does not account for
time-delay, 
reflecting that each node changes its state asynchronously in reality.
We recall the definition of the AsIC model below, in which we consider
choosing a delay-time from the exponential distribution for the sake of
convenience, but of course other distributions such as power-law and
Weibull can be employed.

Let $G = (V, E)$ be a directed graph, where $V$ and $E$ ($\subset V
\times V$) are the sets of all the nodes and the links.  For any $v \in
V$, the set of all the nodes that have links from $v$ is denoted by
$F(v)$ $=$ $\{ u \in V; \ (v, u) \in E \}$ and the set of all the nodes
that have links to $v$ by $B(v)$ $=$ $\{ u \in V; \ (u, v) \in E \}$.
Each node has one of the two states (active and inactive), and the nodes
are called {\em active} if they have been influenced. 
It is assumed that nodes can switch their states only from inactive to
active.

The AsIC model has two types of parameters $p_{u, v}$ and $r_{u,v}$
with $0$ $<$ $p_{u,v}$ $<$ $1$ and $r_{u,v}$ $>$ $0$,
where $p_{u,v}$ and $r_{u,v}$ are referred to as the diffusion
probability through link $(u,v)$ and the time-delay parameter through
link $(u,v)$, respectively. The information diffusion process 
unfolds in continuous-time $t$, and proceeds from a given initial active
node in the following way.  When a node $u$ becomes active at time $t$,
it is given a single chance to activate each currently inactive node $v
\in F(u)$.  A delay-time $\delta$ is chosen from the exponential
distribution with parameter $r_{u,v}$. The node $u$ attempts to activate
the node $v$ if $v$ has not been activated by time $t + \delta$, and
succeeds with probability $p_{u,v}$. If $u$ succeed, $v$ will
become active at time $t + \delta$.  The information diffusion process
terminates if no more activations are possible.

\section{Problem Setting} \label{problem}

We address the {\em hot span detection problem}. 
In this problem, we assume that some change has happened in the way the
information diffuses, and we observe the diffusion sequences of a certain
topic in which the change is embedded, and consider detecting where in
time and how long this change persisted and how big this change is.  We
place a constraint that 
$p_{u, v}$ and $r_{u,v}$ 
do not depend on link $(u, v)$, i.e. $p_{u,v} = p$, $r_{u,v} = r$
($\forall(u, v) \in E$),
which should be acceptable noting that we can naturally assume
that people behave quite similarly when talking about the same topic
(see Section~\ref{discussion}).

Let $[T_1, T_2]$ denote the hot span of the diffusion of a topic, and
let $p_1$ and $p_2$ denote the values of the diffusion probability of
the AsIC model for the normal span and the hot span, respectively.  Note
that $p_1 < p_2$.  A diffusion result of the topic 
is represented as a set of pairs of active nodes and their activation
times; i.e.  $\{(u, t_u), (v, t_v), \cdots \}$.  We consider a diffusion
result $D$ that is generated by the AsIC model with 
$p_1$ for the period $[0, T_1)$, 
$p_2$ for the period $[T_1, T_2]$ and 
$p_1$ for the period $(T_2, \infty)$, where the time-delay parameter
does not change and takes the same value $r$ for the entire period $[0,
\infty)$.  We refer to the set $D$ as a {\em diffusion result with a hot
span}. The problem is reduced to detecting $[T_1, T_2]$ and estimating
$p_1$ and $p_2$ from the observed diffusion results. 
Extensions of this problem setting is discussed later (see
Section\ref{discussion}).

\begin{figure}[tb]
\centering
{\includegraphics[width=5.8cm]{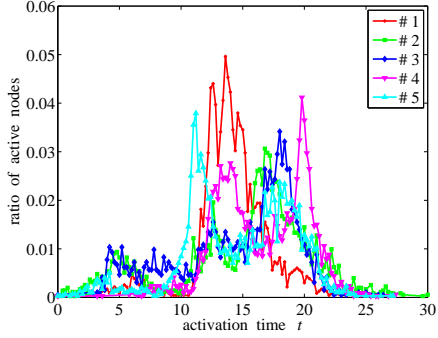}}
\caption{Information diffusion in the blog network with a hot span for 
the AsIC model.}
\label{fig:blog}
\vspace{-0.2cm}
\end{figure}

Figure~\ref{fig:blog} shows examples of diffusion samples with
a hot span based on the AsIC model, where the parameters are set at $p_1
= 0.1$, $p_2 = 0.3$, $r = 1.0$, $T_1 = 10$ and $T_2 = 20$. The network
used is the blog network described later in
Subsection~\ref{datasets}. We plotted the ratio of active nodes (the
number of active nodes at a time step $t$ divided by the number of total
active nodes over the whole time span) for five independent simulations,
each from a randomly chosen initial source node at time $t=0$.
We can clearly see bursty activities around the hot span $[T_1 = 10, T_2
= 20]$.
However, each curve
behaves differently, i.e., some has its bursty activities only in the
first half,
some other has them only in the last half, and yet some other has two
peaks during the hot span. This means that it is quite difficult to
accurately detect the true hot span from only a single diffusion
sample. Methods that use only the observed bursty activities, including
those proposed by
\cite{swan} and 
\cite{kleinberg} would not work. We believe that an explicit use of
underlying diffusion model is essential to solve this problem. It is
crucially important to detect the hot span precisely in order to
identify the external factors which caused the behavioral changes.

\section{Hot Span Detection Methods}\label{methods}

Let $\{D_m; \ m = 1, \cdots, M \}$ be a set of $M$ independent
information diffusion results, 
where $D_m = \{ (u, t_{m,u}), \ (v, t_{m,v}), \
\cdots \}$.  
Each $D_m$ is associated with the observed initial time
$\phi_m = \min \{t_{m,v}; \ (v, t_{m,v}) \in D_m \}$, and the observed
final time $\Phi_m \geq \max \{t_{m,v}; \ (v, t_{m,v}) \in D_m \}$.
We express our observation data by ${\cal D}_M = \{(D_m, \Phi_m); \ m =
1, \cdots, M \}$.  For any $t \in [\phi_m, \Phi_m]$, we set $C_m(t) =
\{v; \ (v, t_{m,v}) \in D_m, \ t_{m,v} < t \}$.  Namely, $C_m(t)$ is the
set of active nodes before time $t$ in the $m$th diffusion result. For
convenience sake, we use $C_m$ as referring to the set of all the active
nodes in the $m$th diffusion result.

\subsection{Parameter Learning Framework}\label{learning}

The following logarithmic likelihood function ${\cal L}({\cal D}_M ; p,
r)$ has been derived to estimate the values of $p$ and $r$ from ${\cal
D}_M$ for the AsIC model in case there is no hot span~\cite{saito:acml09},
\begin{eqnarray}
{\cal L}({\cal D}_M ; p, r) 
\hspace*{-5pt}& = &\hspace*{-5pt} \sum_{m = 1}^{M} {\cal L}((D_m, \Phi_m) ; p, r)\nonumber \\[-0.1cm]
\hspace*{-5pt}& = &\hspace*{-5pt} \sum_{m = 1}^{M} \sum_{v \in C_m} \left( \log h_{m, v} +
      \hspace*{-8pt}\sum_{w \in F(v) \setminus C_m}\hspace*{-5pt} \log g_{m, v, w} \right),
\label{AsIC:objective}
\end{eqnarray}
where $h_{m, v}$ is the probability density that a node $v$ 
$\in D_m$ with $t_{m, v} > 0$ 
is activated at a time $t_{m,v}$, and $g_{m, v, w}$ is the probability
that a node $w$ is not activated by a node $v$ within 
$[\phi_m, \Phi_m]$, where there exists a link $(v, w) \in E$ and $v \in
C_m$.
The values of $p$ and $r$ can be stably
obtained by maximizing Eq.~(\ref{AsIC:objective}) using the EM
algorithm~\cite{saito:acml09}.

The following parameter switching applies for a hot span $S = [T_1,
T_2]$ where ${\cal N}_m$ and ${\cal H}_m$ denote the sets of active
nodes in the $m$-th diffusion result during the normal and the hot
spans, respectively.
\begin{eqnarray*}
p \hspace*{-5pt}&=&\hspace*{-5pt} \left \{ \begin{array}{ll}
\hspace*{-5pt}p_1 &\hspace*{-5pt} 
\mbox{if~$v \in {\cal N}_m(S), \ {\cal N}_m(S) =C_m(T_1) \cup (C_m \setminus C_m(T_2))$}, \\
\hspace*{-5pt}p_2 &\hspace*{-5pt} 
\mbox{if~$v \in {\cal H}_m(S), \ {\cal H}_m(S) = C_m(T_2) \setminus C_m(T_1)$}. \end{array} \right.
\label{parameter-switching}
\end{eqnarray*}


 Then, an extended objective function ${\cal
L}({\cal D}_M ; p_1, p_2, r, S)$ can be defined by adequately modifying
Eq.~(\ref{AsIC:objective}) under this switching scheme.  Clearly,
${\cal L}({\cal D}_M ; p_1, p_2, r, S)$
is expected to be maximized by setting $S$ to the true span $S^* =
[T_1^*, T_2^*]$ if a substantial amount of data ${\cal D}_M$ is
available. 
Thus, our problem is to find the following $\hat{S}$.
\begin{eqnarray}
\hat{S} & = & \arg \max_S {\cal L}({\cal D}_M ; \hat{p}_1, \hat{p}_2, \hat{r}, S),
\label{problem-definition}
\end{eqnarray}
\noindent where $\hat{p}_1$, $\hat{p}_2$, and $\hat{r}$ denote the maximum
likelihood estimators for a given $S$.

In order to obtain $\hat{S}$,
we need to prepare a reasonable set of candidate spans, denoted by
${\cal S}$. One way of doing so is to construct ${\cal S}$ by
considering all pairs of observed activation time points: ${\cal S} =
\{S = [t_1, t_2] : t_1 < t_2, t_1 \in {\cal T}, t_2 \in {\cal T} \}$, where
${\cal T} = \{t_1, \cdots, t_N\}$ is a set of activation time points 
in ${\cal D}_M$.

\subsection{Naive Method}\label{naive}

Now we describe the naive method, which has two iterative loops. In the
inner loop we first obtain the maximum likelihood estimators,
$\hat{p}_1$, $\hat{p}_2$, and $\hat{r}$, for each candidate $S$ by
maximizing ${\cal L}({\cal D}_M ; p_1, p_2, r, S)$ using the EM
algorithm. In the outer loop we select the optimal $\hat{S}$ which gives
the largest ${\cal L}({\cal D}_M ; \hat{p}_1, \hat{p}_2, \hat{r}, S)$
value.  However, this can be extremely inefficient when $N$ is large. To
make it work with a reasonable computational cost, we restrict the
number of candidate time points $N$ to a smaller value $K$ by selecting
$K$ points from ${\cal T}$, i.e., we construct ${\cal S}_K = \{S = [t_1,
t_2] : t_1 < t_2, t_1 \in {\cal T}_k, t_2 \in {\cal T}_K \}$, where
${\cal T}_K = \{t_1, \cdots, t_K\}$. Note that $|{\cal S}_K| =
K(K-1)/2$, which is large when $K$ is large.


\subsection{Proposed Method}\label{proposed}

The naive method should be able to detect the hot span
with a reasonable accuracy when $K$ is set large at the expense of the
computational cost, but the accuracy becomes poorer when $K$
is set smaller to reduce the computational load. We propose a novel detection
method which alleviates this problem and can efficiently and
stably detect a hot span from ${\cal D}_M$.

We first obtain 
$\hat{p}$, and $\hat{r}$, based on the original objective function of
Eq.~(\ref{AsIC:objective}), and focus on 
its first-order derivative with respect to 
$p$ for each node at each individual activation time.  
Let $p_{u, v}$ be the diffusion parameter from a node $u$
to a node $v$. 
The following formula holds for the maximum likelihood estimators due to
the uniform parameter setting of Eq.~(\ref{AsIC:objective}) and the
locally optimal condition.
\begin{eqnarray}
\frac{\partial {\cal L}({\cal D}_M ; \hat{p}, \hat{r})}{\partial p} & = &
\sum_{(u, v) \in E} \frac{\partial {\cal L}({\cal D}_M ; \hat{p}, \hat{r})}{\partial p_{u, v}} \ = \ 0.
\label{AsIC:gradient}
\end{eqnarray}
Consider the following partial sum for a given 
$S = [T_1, T_2]$. 
\begin{eqnarray}
{\cal G}(S) & = &
\sum_{m = 1}^{M} \sum_{(u, v) \in E, u \in {\cal H}_m(S)}
\frac{\partial {\cal L}((D_m, \Phi_m) ; \hat{p}, \hat{r})}{\partial p_{u, v}}.
\label{AsIC:gradient-partial}
\end{eqnarray}


Clearly, ${\cal G}(S)$ 
should be sufficiently large if $S \approx S^*$ due to our problem
setting, which leads to $p_2 > \hat{p} > p_1$. 
Thus, the hot span $S^*$ can be estimated by searching for ${\hat S}$
that maximizes ${\cal G}(S)$.
\begin{eqnarray}
{\hat S} & = & \arg \max_{S \in {\cal S}} {\cal G}(S).
\label{detection-method}
\end{eqnarray}

The nice thing here is that we can incrementally calculate ${\cal G}(S)$ by
Eq.~(\ref{detection-method2}),
where ${\cal T} = \{t_1, \cdots, t_N\}$ and $t_i < t_j$ if $i < j$.

\vspace{-0.2cm}

\begin{eqnarray}
\hspace*{-12pt}
{\cal G}([t_i, t_{j+1}]) \hspace*{-5pt}& = &\hspace*{-5pt} {\cal G}([t_i, t_j]) +
\sum_{m = 1}^{M} \hspace*{-10pt} \sum_{\stackrel{\scriptstyle (u, v) \in E}{u \in C_m(t_{j+1}) \setminus C_m(t_j)}}
\hspace*{-17pt}\frac{\partial {\cal L}((D_m, \Phi_m) ; \hat{p}, \hat{r})}{\partial p_{u, v}}.
\label{detection-method2}
\end{eqnarray}


The computational cost 
for examining each candidate span is much smaller than the naive method
described above.  Thus, we can use all the pairs to construct 
${\cal S}$. We summarize our proposed method below.
\begin{description}
\item[1.] Maximize ${\cal L}({\cal D}_M ; p, r)$ by using the EM algorithm.
\item[2.] Construct 
${\cal T}$ and
${\cal S}$.
\item[3.] Detect 
$\hat{S}$ by Eq.~(\ref{detection-method}) and output $\hat{S}$.
\item[4.] Maximize ${\cal L}({\cal D}_M ; p_1, p_2, r, \hat{S})$ by using the EM algorithm,
and output $\hat{p}_1$, $\hat{p}_2$, and $\hat{r}$.
\end{description}
Here note that the proposed method requires maximization by using the EM
algorithm only twice.

\section{Experiments}\label{experiments}

We experimentally investigated how accurately the proposed method
can estimate both the hot span and the diffusion probabilities
in the hot and normal spans, as well as its efficiency,
by comparing it with the naive method using three real world networks.
We used three different values for $K$, {\it i.e.}, $K=5$, $10$, and
$20$ for the naive method.

The derivation assumed that there are multiple observed data
sequences, but in the experiments we chose to learn from a single
sequence, {\it i.e.}, $M=1$, which is the most difficult situation.

\subsection{Datasets}\label{datasets}

The three data are all bidirectionally connected networks.
The first one is a trackback network of Japanese blogs used in 
\cite{kimura:tkdd}, which has $12,047$ nodes and $79,920$ directed 
links (the blog network).
The second one is a coauthorship network used in \cite{palla},
which has $12,357$ nodes and $38,896$ directed links (the Coauthorship 
network).
The last one is a network of people that was derived from the 
``list of people'' within Japanese Wikipedia, used in 
\cite{kimura:tkdd}, and has $9,481$ nodes and $245,044$ directed 
links (the Wikipedia network). 

For these networks, we generated diffusion samples with a hot span using
the AsIC model.
According to \cite{kempe:kdd}, we set the diffusion probability for the
normal span, $p_1$, to be a value smaller than $1/{\bar d}$, where
$\bar{d}$ is the mean out-degree of a network, and set the diffusion
probability for the hot span, $p_2$,  to be three times
larger than $p_1$. Thus, $p_1$ and $p_2$ are $0.1$ and $0.3$ for the
blog network, $0.2$ and $0.6$ for the Coauthorship network, and $0.02$
and $0.06$ for the Wikipedia network, respectively.  We fixed the
time-delay parameter at 1 ($r =1$) for all the networks because changing
$r$ works only for scaling the time axis of the diffusion results.  We
set the hot span to $[T_1 = 10, T_2 = 20]$ based on the observation on
the preliminary experiments. In all we generated five information
diffusion samples using these parameter values for each network,
randomly selecting an initial active node for each diffusion sample.

\subsection{Results}\label{results}

\begin{figure*}[!tb]
\centering
\subfloat[Blog\label{blog_span}]
{\includegraphics[width=3.7cm]{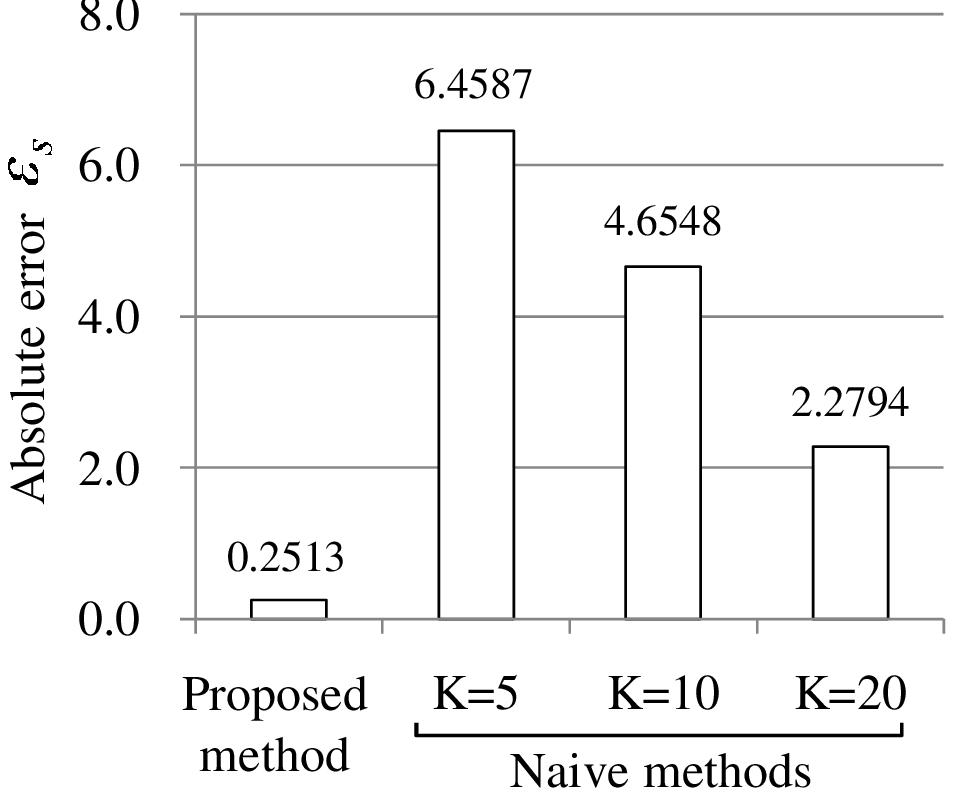}}
\hspace{15mm}
\subfloat[Coauthorship\label{coauthor_span}]
{\includegraphics[width=3.7cm]{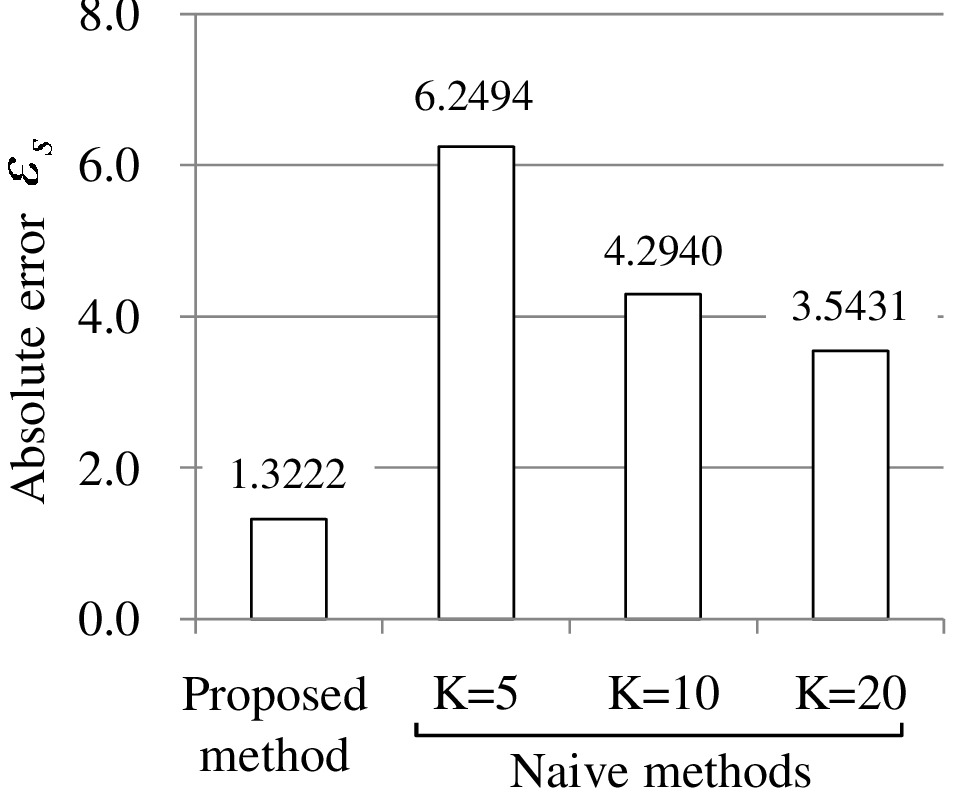}}
\hspace{15mm}
\subfloat[Wikipedia\label{wiki_span}]
{\includegraphics[width=3.7cm]{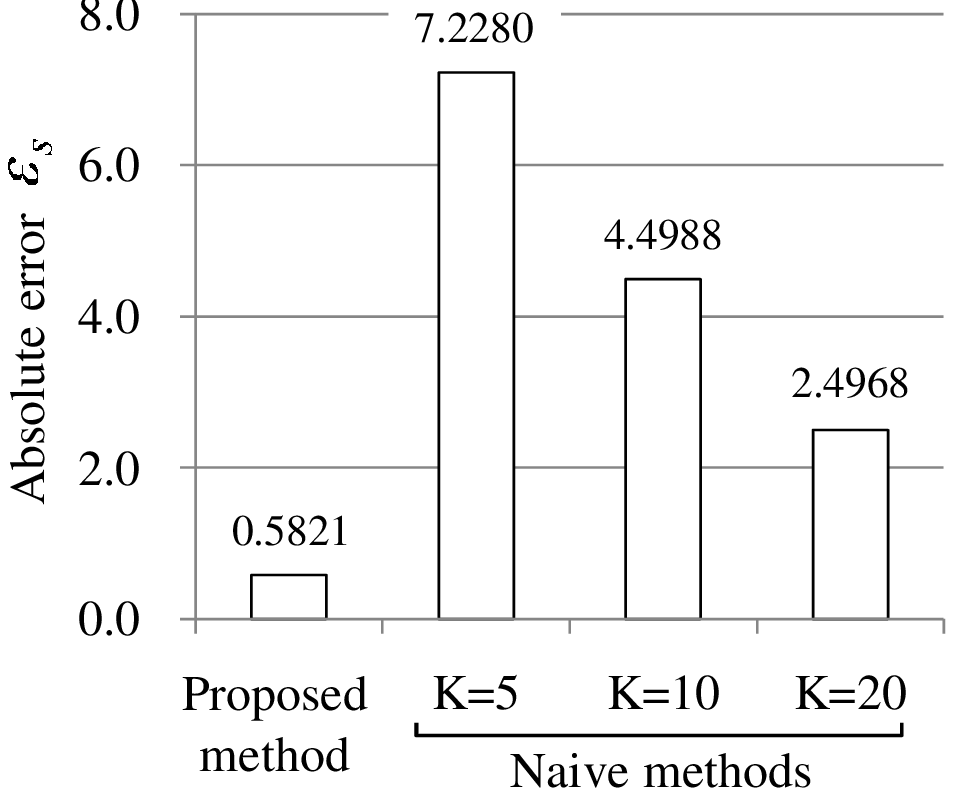}}\\
\vspace{-0.3cm}
\caption{Comparison in accuracies of the estimated hot span}
\label{fig:span}
\subfloat[Blog\label{blog_prob}]
{\includegraphics[width=3.7cm]{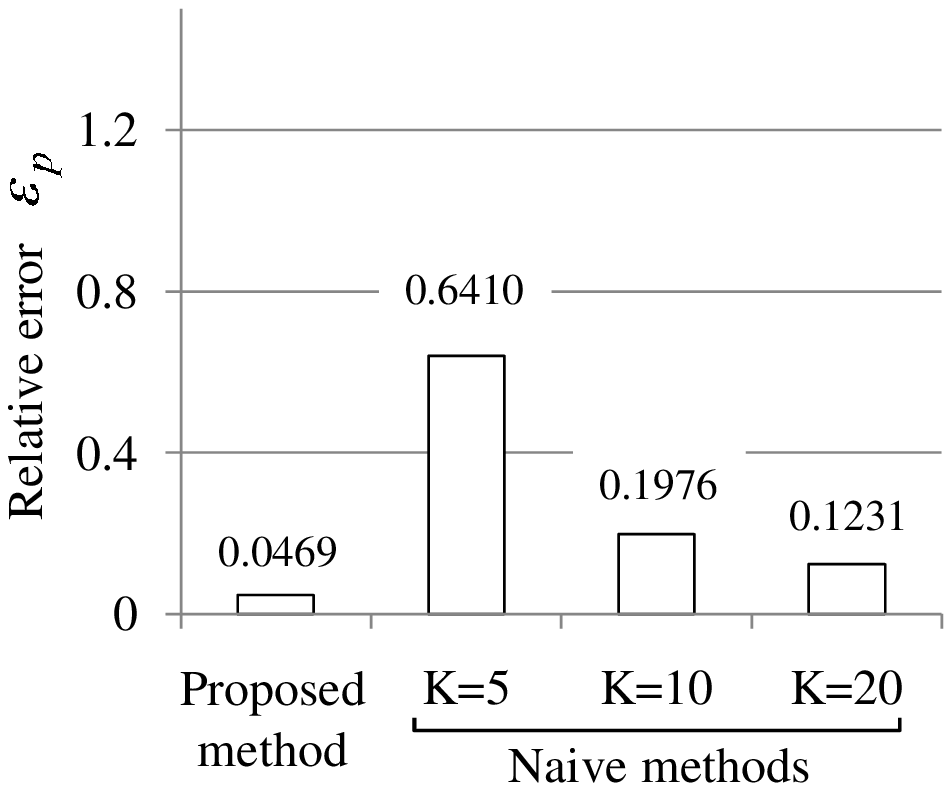}}
\hspace{15mm}
\subfloat[Coauthorship\label{coauthor_prob}]
{\includegraphics[width=3.7cm]{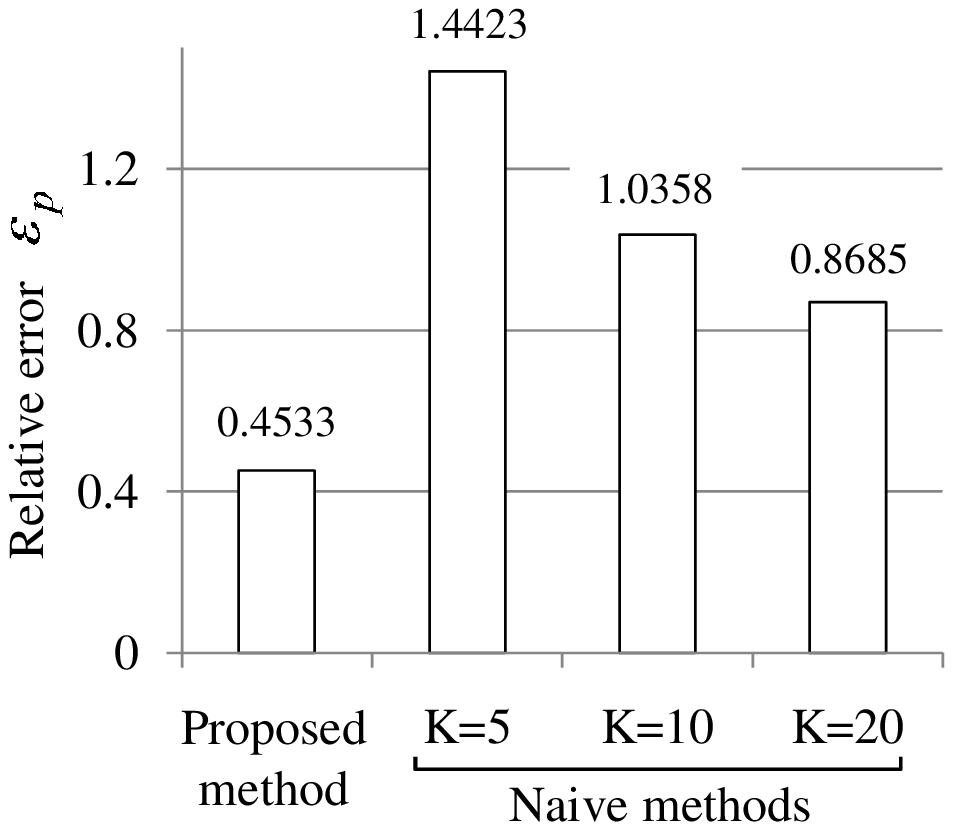}}
\hspace{15mm}
\subfloat[Wikipedia\label{wiki_prob}]
{\includegraphics[width=3.7cm]{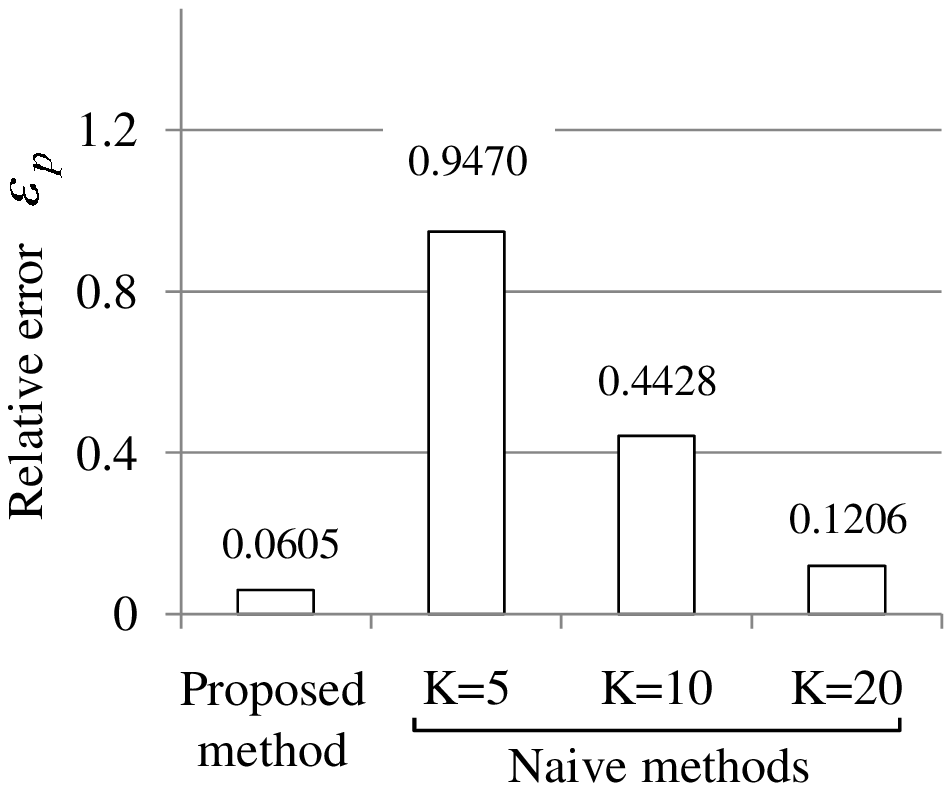}}\\
\vspace{-0.3cm}
\caption{Comparison in accuracies of the estimated diffusion probability}
\label{fig:prob}
\subfloat[Blog\label{blog_time}]
{\includegraphics[width=3.7cm]{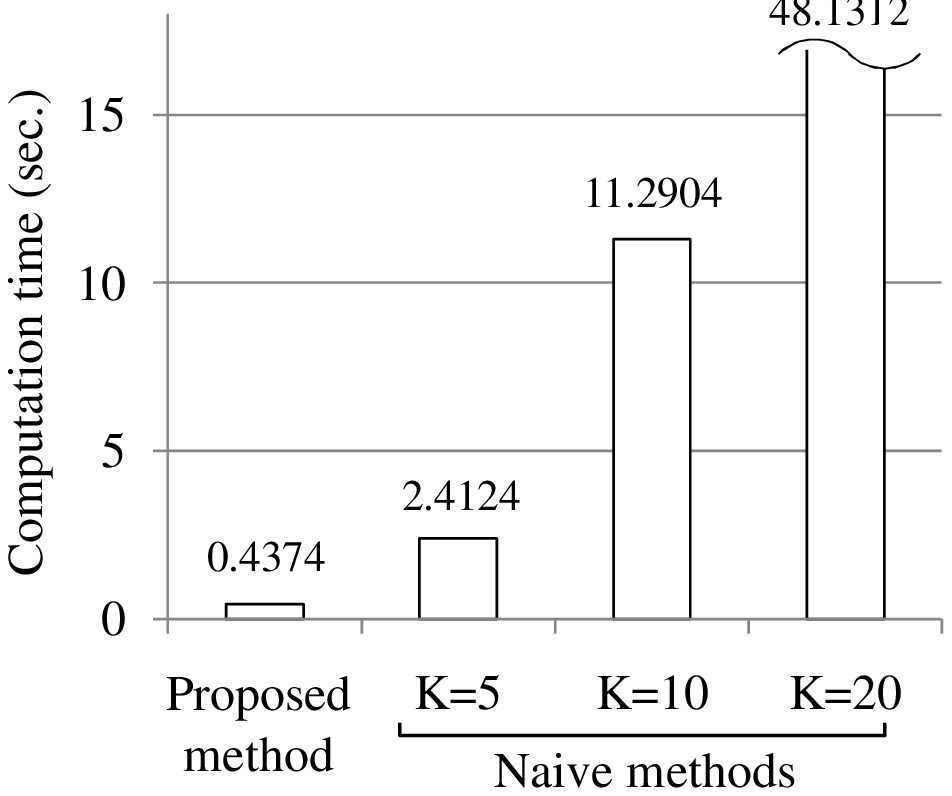}}
\hspace{15mm}
\subfloat[Coauthorship\label{coauthor_time}]
{\includegraphics[width=3.7cm]{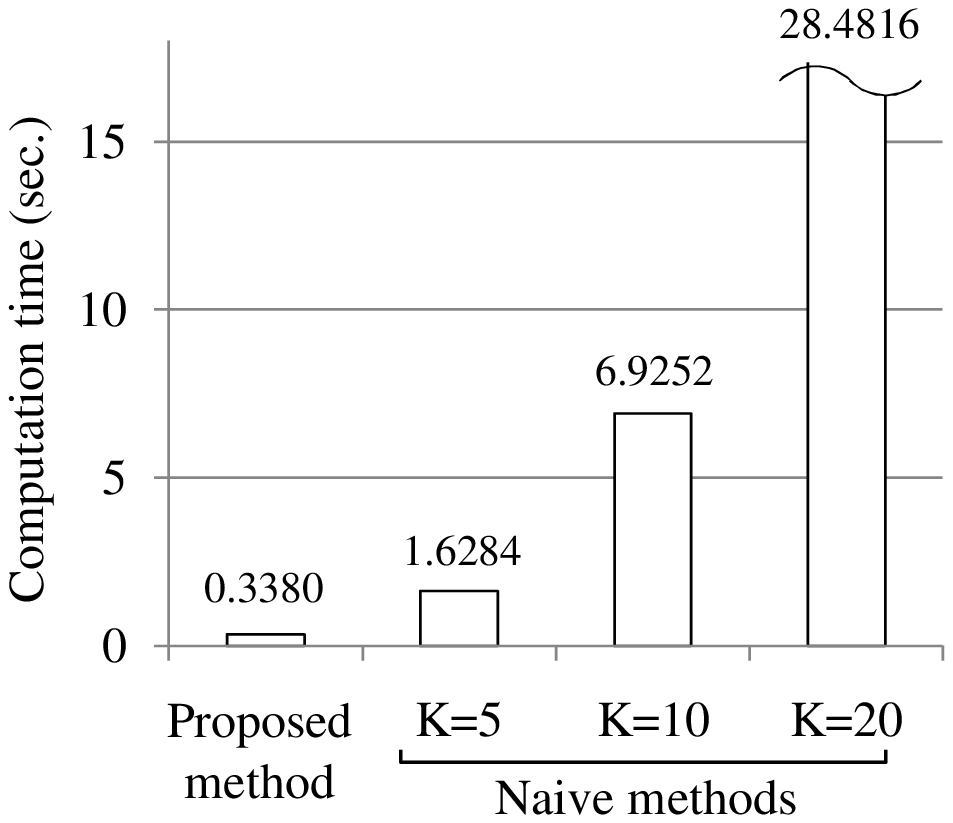}}
\hspace{15mm}
\subfloat[Wikipedia\label{wiki_time}]
{\includegraphics[width=3.7cm]{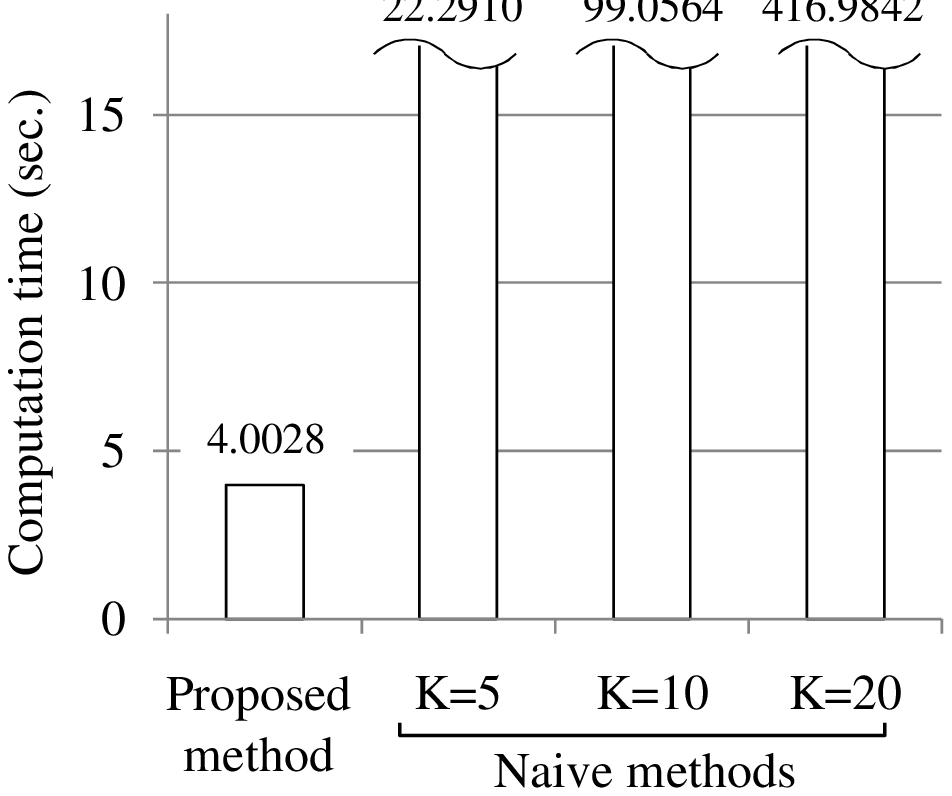}}
\vspace{-0.3cm}
\caption{Comparison in computation time}
\label{fig:time}
\vspace{-0.4cm}
\end{figure*}


We compared the proposed method with the naive method in terms of 1) the
accuracy of the estimated hot span ${\hat{\cal S}}=[\hat{T}_1,
\hat{T}_2]$, 2) the accuracy of the diffusion probabilities $p_1$ (for
the normal span) and $p_2$ (for the hot span), and 3) the computation
time. Both the proposed and the naive methods were tested to each
diffusion sample mentioned above, and the results were averaged over the
five independent trials for each network.

Figure \ref{fig:span} shows the accuracy for ${\hat{\cal S}}$ in
the absolute error ${\cal E}_s = | \hat{T}_1 - T_1 | + | \hat{T}_2 - T_2
|$.  We see that the proposed method achieves a good accuracy, much
better than the naive method for every network. As expected, ${\cal
E}_s$ for the naive method decreases as $K$ becomes larger. But, even in
the best case ($K=20$), its average error is about 3 to 10 times larger
than that of the proposed method.
Figure~\ref{fig:prob} shows the accuracy of $p_1$ and $p_2$
in the relative error ${\cal E}_p = | \hat{p}_1 - p_1 | / p_1
+ | \hat{p}_2 - p_2 | / p_2$.
Here again, the average relative error for the naive method decreases as $K$ becomes
larger. However, even in the best case ($K=20$), it is
about 2 to 3 times larger than that of the proposed method. 
We note that the average errors for the Coauthorship network are
relatively large. This is because the number of active nodes within the
normal span was relatively small for this network.
Figure~\ref{fig:time} shows the computation time.
It is clear that the proposed method is much faster than the naive
method.
The significant difference is attributed to the difference in
the number of runs of the EM algorithm. The proposed method executes the
EM algorithm only twice: 
steps 1 and 4 in the algorithm (see Section \ref{proposed}).  On the
other hand, the naive method has to execute the EM algorithm once for
every single candidate span $S \in {\cal S}_K$
which is $|{\cal S}_K| = K (K - 1)/2$ times (see Section \ref{naive}).
Indeed, the computation time of the naive method for $K=5$ is about 5
times larger 
for every network, which
is consistent with 
$|{\cal S}_K| = 10$.
This relation roughly holds also for the other two cases ($K=10$ and
$K=20$). This means that even if the naive method could achieve a good
accuracy by setting $K$ to a sufficiently large value, it would require
unacceptable computation time for such a large $K$.

In summary, we can say that the proposed method can detect and estimate
the hot span and diffusion probabilities much more accurately and
efficiently compared with the naive method. Here we mention that
we could obtain much better results by using more than one diffusion
sequence, say $M = 5$, but we have to omit the details due to space
limitations.

\section{Discussion}\label{discussion}

We placed a simplifying constraint that the parameters $p_{u, v}$ and
$r_{u,v}$ are link independent, i.e. $p_{u,v} = p$, $r_{u,v} = r$
($\forall(u, v) \in E$), by focusing on single topic diffusion
sequences.  \cite{saito:acml09,
saito:ecml10} gave some
evidences for this assumption. They examined $7,356$ diffusion sequences
for a real blogroll network containing $52,525$ bloggers and $115,552$
blogroll links, and experimentally confirmed that $p$ and $r$ that were
learned from different diffusion sequences belonging to the same topic
were quite similar for most of the topics. This observation naturally
suggests that people behave quite similarly for the same topic.

In this paper, we considered AsIC model, but it is straightforward to
apply the same technique to AsLT model~\cite{saito:ecml10} and to their SIS
versions in which each node is allowed to be activated multiple
times. The same idea can naturally be applied to opinion formation
model, e.g. value-weighted voter model~\cite{kimura:aaai10}.

\begin{figure}[tb]
\centering
{\includegraphics[width=6cm]{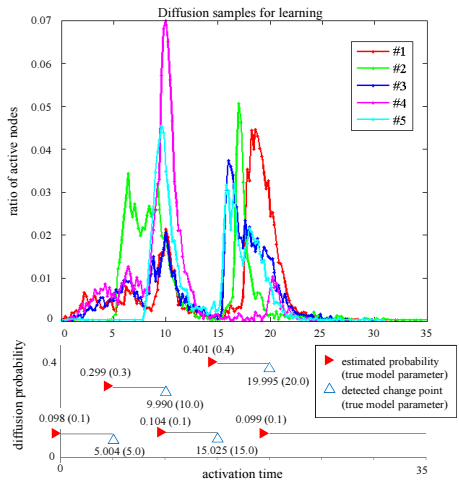}}
\caption{Information diffusion in the blog network with two hot spans for 
the AsIC model.}
\label{fig:two-peak}
\vspace{-0.2cm}
\end{figure}

The change pattern considered here is the simplest one. We can assume a
more intricate problem setting such that both $p$ and $r$ change for
multiple distinct hot spans and the shape of change pattern $p$ is not
necessarily rect-linear. One possible extension is to approximate the
pattern of any shape by $J$ pairs of time interval each with its
corresponding $p_j$, {\it i.e.}, $Z_J = \{ ([t_{j-1}, t_j],p_j);\ j = 1,
\cdots J\}\  (t_0=0, t_J=\infty)$ and use a divide-and-conquer type greedy
recursive partitioning, still employing the derivative of the likelihood
function ${\cal G}$ as the main measure for search.  More specifically,
we first initialize $Z_1 = \{ ([0, \infty), \hat{p}_1)\}$ where
$\hat{p}_1$ is the maximum likelihood estimator, and search for the
first change time point $t_1$, which we expect to be the most
distinguished one, by maximizing $|{\cal G}([t,
\infty),\hat{p}_1)|$.\footnote{Note that the total sum of ${\cal
G}=0$.} We recursively perform this operation $J$ times by fixing the
previously determined change points. When to stop can be determined by a
statistical criterion such as AIC or MDL. This algorithm requires
parameter optimization $J$ times.  Figure~\ref{fig:two-peak} is one of
the preliminary results obtained for two distinct rect-linear patterns
using five sequences ($M=5$) in case of the blog network. MDL is used as
the stopping criterion. The change pattern of $p$ is almost perfectly
detected with respect to both $p_j$ and $t_j$ ($J=5$).

\section{Conclusion}\label{conclusion}

In this paper, we addressed the problem of detecting the change in
behavior of information diffusion from a limited amount of observed
diffusion sequences in a retrospective setting, assuming that the
diffusion follows the asynchronous independent cascade (AsIC) model.  We
defined the ``hot span'' as the period during which the diffusion
probability is changed to a relatively high value compared with the
other periods (called the normal spans). A naive method to detect such a
hot span would have to iteratively update the candidate hot span
boundaries, each requiring parameter optimization such that the
likelihood function is maximized. This is very inefficient and totally
unacceptable. We developed a novel and general framework that avoids the
inner loop optimization during search by making use of the first
derivative of the likelihood function. It needs to optimize the
parameter values only twice by the iterative updating algorithm (EM
algorithm), which reduces the computation times by 5 to 100 times, and
is very efficient. We compared the proposed method with the naive method
that considers only the randomly selected boundary candidates, by
applying both the methods (the proposed and the naive) to information
diffusion samples generated by simulation from three real world large
networks, and confirmed that the proposed method far outperforms the
naive method both in terms of accuracy and efficiency.  Although we
assumed a very simplified problem setting in this paper, the proposed
method can be easily extended to solve more intricate problems. We
showed one possible direction and the preliminary results obtained for
two rect-linear shape hot spans was very promising. Our immediate
future work is 
to evaluate our method using real world information diffusion samples 
with hot spans, as well as 
to deal with spatio-temporal hot span detection problems using more 
appropriate stochastic models under a similar problem solving framework. 

\section*{Acknowledgments}

This work was partly supported by Asian Office of Aerospace Research and
Development, Air Force Office of Scientific Research under Grant
No. AOARD-10-4053, and JSPS Grant-in-Aid for Scientific Research (C)
(No. 23500194).


\bibliographystyle{named}
\bibliography{ourbib}

\end{document}